\documentclass[reprint,showpacs,aps,pra]{revtex4-1}
\usepackage{graphics}
\usepackage{bm}
\usepackage{amsmath}
\usepackage{amssymb}
\usepackage{graphicx}
\usepackage{amsthm}
\begin{document}

\title{Entanglement  in fermion systems}
\author{N. Gigena, R.Rossignoli} \affiliation{Departamento de F\'{\i}sica-IFLP,
Universidad Nacional de La Plata, C.C. 67, La Plata (1900), Argentina}


\begin{abstract}We analyze the problem of quantifying entanglement in pure
and mixed states of fermionic systems with fixed number parity yet not
necessarily fixed particle number. The ``mode entanglement'' between one
single-particle level and its orthogonal complement is first considered, and an
entanglement entropy for such a partition of a particular basis of the
single-particle Hilbert space $\mathcal H$ is defined. The sum over all
single-particle modes of this entropy is introduced as a measure of the total
entanglement of the system with respect to the chosen basis and it is shown
that its minimum over all bases of $\mathcal H$ is a function of the one-body
density matrix. Furthermore, we show that if minimization is extended to all
bases related through a Bogoliubov transformation, then the entanglement
entropy is a function of the generalized one-body density matrix. These results
are then used to quantify entanglement in fermion systems with four
single-particle levels. For general pure states of such a system a closed
expression for the fermionic concurrence is derived, which generalizes  the
Slater correlation measure defined in [J. Schliemann et al, Phys.\ Rev.\ A {\bf
64}, 022303 (2001)], implying that ``particle entanglement'' may be seen as
minimum ``mode entanglement''. It is also shown that the entanglement entropy
defined before is related to this concurrence by an expression analogous to
that of the two-qubit case. For  mixed states of this system the convex roof
extension of the previous concurrence and entanglement entropy are evaluated
analytically, extending the results of previous ref.\ to general states.
\end{abstract}
 \maketitle
\section{Introduction}
Quantum entanglement is not only one of the key features of quantum mechanics
but also an essential resource in quantum information processing \cite{NC.00}.
It plays a central role in quantum teleportation \cite{BB.93} and  quantum
computation \cite{JL.03}. Consequently, the understanding and quantification of
this resource has become a fundamental problem in quantum information theory
\cite{HH.09}. It has also provided deep insights into the structure of
correlations and quantum phase transitions in many-body systems
\cite{VL.03,AF.08,ECP.10}.

If $|\Psi_{AB}\rangle\in\mathcal{H}_A\otimes\mathcal{H}_B$ is a pure state of a
composite quantum system, its entanglement is quantified by the entanglement
entropy $S(\rho_A)=S({\rm Tr_B}|\Psi_{AB}\rangle\langle \Psi_{AB}|)=S(\rho_B)$,
where $S(\rho)=-{\rm Tr}\rho\log_2\rho$ is the von Neumann entropy. It is then
seen that the notion of entanglement in such systems relies on the tensor
product structure of its state-space \cite{Pe.95}. In fermionic systems,
however, the situation is less clear since the state-space has no longer this
structure due to indistinguishability.

When generalizing the notion of entanglement to systems of indistinguishable
particles
\cite{Sch.01,SDM.01,Eck.02,Za.02,Shi.03,Fri.13,Wi.03,GM.04,Pl.09,IV.13} mainly
two different approaches have been taken: \emph{Mode entanglement}
\cite{Za.02,Shi.03,Fri.13,Be.14,Pu.14} and \emph{quantum
correlations}/\emph{particle entanglement}
\cite{Sch.01,SDM.01,Eck.02,Wi.03,GM.04,Pl.09,IV.13,Ci.09,Os.14,SL.14}. In the
first case the parties share different modes of a given basis of the
single-particle Hilbert space. Therefore, mode entanglement of a system does
not remain invariant with respect to unitary transformations in the
single-particle (sp) space. The second approach looks for correlations between
particles and beyond antisymmetrization.  In \cite{Sch.01,Eck.02} a fermionic
analog of the Schmidt decomposition and Schmidt number was introduced to
quantify entanglement in two-fermion systems, and also a fermionic
``concurrence'' was defined. While these measures of entanglement remain
invariant under unitary transformations in the sp space, they are restricted to
states with a fixed particle number, which is not the general case in fermionic
systems. The same problem arises in \cite{Wi.03}, where in order to share
particles between parties it is necessary to project the original state onto
subspaces with definite particle number.

In this paper we first consider pure states of fermionic systems within a
grand-canonical context, so the particle number is not necessarily fixed.
Fermionic states with no fixed number of fermions arise, for instance, when
considering the vacuum of quasiparticles defined through a Bogoliubov
transformation \cite{RS.80,Ci.09,Pu.14}, as well as by simply applying
particle-hole transformations, such that the state is viewed as a vacuum of
certain fermion operators plus particle-hole excitations. The fermion number
parity of these states is nonetheless fixed, in agreement with fermionic
super-selection rules \cite{Fr.15}. The entanglement between a single fermionic
mode and the remaining sp orthogonal space in such states is first considered,
and an entanglement entropy is defined in order to quantify these correlations.
We then propose the sum over single-particle modes of this entropy as a measure
of the total mode entanglement associated with the chosen sp basis, and show
that its minimum over all single particle bases depends only on the eigenvalues
of the one-body density matrix $\rho^{\rm sp}_{ij}=\langle c_j^\dagger c_i
\rangle$, being therefore invariant under sp transformations. Furthermore, it
is  shown that if the minimization is extended to all quasiparticle bases,
i.e., bases related through Bogoliubov transformations, the minimum
entanglement entropy is just the von Neumann entropy of the generalized
one-body density matrix $\rho^{\rm qsp}$, which contains in addition the pair
creation and annihilation contractions $\langle c^\dagger_j c^\dagger_i\rangle$
and $\langle c_j c_i\rangle$. Its convex roof extension for mixed states is
also introduced. This quantity allows to rigorously identify mixed states which
cannot be written as convex mixtures of Slater determinants or quasiparticle
vacua (or in general fermionic gaussian states \cite{Ci.09,Os.14}), like
thermal states of interacting fermion systems at sufficiently low temperatures,
quantifying their quantum correlations.

We then focus on fermionic systems with sp space dimension 4. For general
states it is shown that the minimum over all quasiparticle bases of the
entanglement entropy can be written in terms of a fermionic analog of the
concurrence \cite{Os.14,SL.14,W.98}, that reduces to the Slater correlation
measure defined in \cite{Sch.01,Eck.02} for two-fermion states. Its convex roof
extension for mixed states is also evaluated analytically, extending explicitly
the results of \cite{Sch.01} to arbitrary mixed states with fixed number parity
\cite{Os.14}. This allows to evaluate in closed form the convex-roof extension
of the previous entanglement entropy. A simple illustrative example is
provided.

\section{Formalism}
\subsection{Single level entanglement entropy}
We start by considering a pure state $|\Psi\rangle$ of a fermion system with an
$n$-dimensional single-particle (sp)  Hilbert space $\mathcal{H}$. The system
is described by a set of fermion annihilation and creation operators $\{c_j,
c^\dagger_j,\}$ satisfying
\begin{equation}c_ic_j+c_jc_i=0,\;\;c_i c_j^\dagger+c^\dagger_j c_i=\delta_{ij}
\,,\label{1}
 \end{equation}
such that $\{c^\dagger_j|0\rangle,\;j=1,\ldots,n\}$ is an orthonormal basis of
sp states ($|0\rangle$ denotes the vacuum of the operators $c_j$). We will
work within a general grand-canonical context, in which $|\Psi\rangle$ is not
necessarily a state with a definite value of the fermion number $N=\sum_j c^\dagger_j c_j$.
It may be, for instance, a vacuum of qusiparticle operators $a_\nu$, related
with the $c_j$'s through a Bogoliubov transformation \cite{RS.80}. In this
case, it is a sum of pure states with different fermion numbers (see Appendix),
yet having all the same number parity
\begin{equation}P=\exp[i\pi \sum_j c^\dagger_j c_j]\label{2}\,, \end{equation}
such that $P|\Psi\rangle=\pm|\Psi\rangle$. Let us also recall that the
elementary particle-hole Bogoliubov transformation
\begin{equation}c_j\rightarrow c_j^\dagger, \;\;c_j^\dagger\rightarrow c_j\,,
\label{3} \end{equation}
leaves the anticommutation relations unchanged, so that formally, it is a
matter of choice whether one considers the particles or the holes as the
``true'' fermions. We will take this basic symmetry into account in all the
following correlation measures, such that they all remain invariant under the
previous transformation. We will just assume that all pure states involved have
a definite number parity \cite{Fr.15}, which implies  $\langle c_j\rangle=0$
and also  $\langle O\rangle=0$ for any operator $O$ which is a product of an
odd number of fermion operators $c_j$, $c^\dagger_j$.

We now consider a partition $(A,B)$ of $\mathcal{H}$, where $A$ denotes the
single mode or ``level'' $j$ and $B$ the remaining orthogonal sp space. Eq.\
(\ref{1}) implies that the operators
\begin{equation}\Pi_j=c^\dagger_j c_j,\;\;\Pi_{\bar{j}}=c_j c^\dagger_j\,,\;\;\Pi_j
 +\Pi_{\bar{j}}=1\label{4},  \end{equation}
constitute a basic set of orthogonal projectors, defining a standard projective
measurement on the level $j$. Accordingly, we may decompose  any state
$|\Psi\rangle$ as
\begin{eqnarray}
|\Psi\rangle&=&c^\dagger_j c_j|\Psi\rangle+c_jc^\dagger_j|\Psi\rangle\label{psi1}\\
&=&\sqrt{p_j}\,|\Psi_j\rangle+\sqrt{p_{\bar{j}}}\,|\Psi_{\bar{j}}\,,\rangle\label{psi2}
\end{eqnarray}
where the first (second) term in (\ref{psi1}) selects  the component of
$|\Psi\rangle$ where the state $j$ is occupied (empty) and
$|\Psi_j\rangle=\frac{1}{\sqrt{p}_j}c^\dagger_j c_j|\Psi\rangle$,
$|\Psi_{\bar{j}}\rangle=\frac{1}{\sqrt{p_{\bar{j}}}}
c_jc^\dagger_j|\Psi\rangle$ are the corresponding normalized states. Here $p_j$
($p_{\bar{j}}$) is the probability of finding the level $j$ occupied (empty) in
$|\Psi\rangle$:
 \begin{equation} p_j=\langle \Psi|c^\dagger_j c_j|\Psi\rangle\,,\;\;p_{\bar{j}}
 =\langle \Psi|c_jc^\dagger_j|\Psi\rangle=1-p_j\,.\end{equation}
For an operator $O_A$ depending just on $c_j, c^\dagger_j$ and $O_B$ depending
just on the complementary set  $\{c_k, c^\dagger_k, k\neq j\}$, we then obtain,
assuming $P|\Psi\rangle=\pm |\Psi\rangle$,
  \begin{eqnarray}
    \langle\Psi|O_{A(B)}|\Psi\rangle
    &=&p_j\langle\Psi_j|O_{A(B)}|\Psi_j\rangle
  +p_{\bar{j}}\langle \Psi_{\bar{j}}|O_{A(B)}|\Psi_{\bar{j}}\rangle\nonumber\\
    &=&{\rm tr}_{A(B)}\rho_{A(B)}O_{A(B)}\,,
    \end{eqnarray}
where $\rho_A=p_jc^\dagger_j|0\rangle\langle 0|c_j+p_{\bar{j}}|0\rangle\langle
0|$ and $\rho_B=p_jc_j|\Psi_j\rangle\langle \Psi_j|c_j^\dagger+
p_{\bar{j}}|\Psi_{\bar{j}}\rangle\langle\Psi_{\bar{j}}|$ represent reduced
density operators for systems $A$ and $B$ respectively.

The entanglement between $A$ and $B$ can then be quantified by the entropy of
the elementary distribution $\{p_j,p_{\bar{j}}=1-p_j\}$:
\begin{eqnarray}
 S(\rho_A)=S(\rho_B)&=&-p_j\log_2 p_j-(1-p_j)\log_2(1-p_j)\,,\label{S1}\\
 &=&h(p_j)\,,\end{eqnarray}
where $S(\rho)=-{\rm Tr}\rho\log_2\rho$ is the von Neumann entropy and
$h(p)=-p\log_2 p-(1-p)\log_2(p)$ ($0\leq h(p)\leq 1$). This entropy remains
obviously invariant after a particle-hole transformation (\ref{3}). For a pure
state $|\Psi\rangle$, Eq.\ (\ref{S1}) vanishes {\it if and only if} (iff)
$|\Psi\rangle$ is separable with respect to this level, i.e., iff the level $j$
is either occupied ($p_j=1$) or empty ($p_j=0$) in $|\Psi\rangle$, such that
$|\Psi\rangle=c^\dagger_j c_j|\Psi\rangle$ or
$|\Psi\rangle=c_jc^\dagger_j|\Psi\rangle$ respectively. Its maximum value $1$
is attained for $p_j=1/2$.

\subsection{One-body entanglement entropy}
The sum
\begin{equation}
S_{\bm c}=\sum_j h(p_j),\label{5}
\end{equation}
is a measure of the entanglement associated with the sp basis determined by the
operators $c^\dagger_j$. Eq.\ (\ref{5}) vanishes iff each level $j$ of this
basis is disentangled from its complementary sp space, i.e., iff each level is
either occupied or empty in $|\Psi\rangle$, such that $|\Psi\rangle$ is a
Slater determinant in this basis: $|\Psi\rangle=c^\dagger_{j_1}\ldots
c^\dagger_{j_m}|0\rangle$ for some subset of levels $\{j_1,\ldots,j_m\}$.

Eq.\ (\ref{5}) depends on the choice of sp basis, i.e., on the choice of
fermion operators $\bm{c}=(c_1\ldots,c_n)^T$. We now consider the minimum of
(\ref{5}) over all  sp bases of ${\cal H}$, i.e.,
\begin{equation} S^{\rm sp}=\mathop{\rm Min}_{\bm{c}'}S_{\bm{c}'}\,,
\label{6}\end{equation}
where  $S_{\bm{c}'}=\sum_j h(p'_j)$, with $p'_j=\langle \Psi|{c'}^\dagger_j
c'_j|\Psi\rangle$ and $\bm{c'}=(c'_1,\ldots, c'_n)^T$ an arbitrary  set of
fermion operators related with the $c_j$'s  through a unitary transformation:
\begin{equation} \bm{c}'=U^\dagger \bm{c}\label{uca}\,, \end{equation}
with $U$  a $n\times n$ unitary matrix (such that the fermionic relations
(\ref{1}) are preserved). Eq.\ (\ref{6}) vanishes {\it iff $|\Psi\rangle$ is a
Slater determinant}, i.e., $|\Psi\rangle={c'}^\dagger_{k_1}\ldots
{c'}^\dagger_{k_m}|0\rangle$ for some operators $c'_k$ of the form (\ref{uca}).
Hence, $S^{\rm sp}=0$ iff there is a sp basis where every level is disentangled
from its complementary sp space.

Defining the sp density matrix $\rho^{\rm sp}=1-\langle
\bm{c}\bm{c}^\dagger\rangle$ (with $\langle
O\rangle\equiv\langle\Psi|O|\Psi\rangle$),  of elements
\begin{equation}
\rho_{ij}^{\rm sp}=\langle c^\dagger_{j}c_i\rangle\,,\label{rhoc}
\end{equation}
it is seen that the minimum (\ref{6}) is  reached for those operators
$\bm{c'}$ which {\it diagonalize} $\rho^{\rm sp}$, i.e.,  satisfying
\begin{equation} \langle {c'}^\dagger_k c'_l\rangle=
(U^\dagger \rho^{\rm sp} U)_{lk}=p'_k\delta_{kl}\label{ad}\,,  \end{equation}
 with $p'_k$ the eigenvalues of $\rho^{\rm sp}$. \\
{\it Proof:} Eqs.\ (\ref{uca})--(\ref{ad}) imply that $p_j=\rho^{\rm
sp}_{jj}=\sum_k |U_{jk}^2|p'_k$. Hence,  concavity of the function $h(p)$
entails $\sum_j h(p_j)\geq \sum_{j,k}|U_{jk}^2|h(p'_k)=\sum_k h(p'_k)$, with
equality reached iff the $p_j$'s are already the eigenvalues of $\rho^{\rm
sp}$.  \qed

The minimum value (\ref{6}) can then be expressed as
\begin{equation}S^{\rm sp}=\sum_k h(p'_k)={\rm tr}\,h(\rho^{\rm sp})\,,
 \label{66}\end{equation}
being now apparent that $S^{\rm sp}$ vanishes iff the eigenvalues $p'_k$  are
either $0$ or $1$, i.e., iff $(\rho^{\rm sp})^2=\rho^{\rm sp}$, a condition
ensuring that $|\Psi\rangle$ is a Slater determinant \cite{RS.80}.

Eq.\ (\ref{66}) has in addition  the obvious meaning of quantifying how mixed
or ``hot'' is $|\Psi\rangle$  with respect to the set of all one-body operators
of the form
\begin{equation} O=\sum_{i,j} o_{ij}c^\dagger_i c_j\,,\label{O}\end{equation}
since their  averages are completely determined just by $\rho^{\rm sp}$:
$\langle \Psi| O|\Psi\rangle={\rm tr}\,\rho^{\rm sp} o$. Accordingly, $S^{\rm
sp}$ remains invariant under one-body unitary transformations
$|\Psi\rangle\rightarrow \exp(-iO)|\Psi\rangle$, with $O$  any hermitian
one-body operator of form (\ref{O}), since they lead to a unitary
transformation of $\rho^{\rm sp}$ ($\rho^{\rm sp}\rightarrow U\rho^{\rm sp}
U^\dagger$, with $U=e^{-i o}$) and hence do not affect its eigenvalues.

\subsection{Generalized one-body entanglement entropy}
A quasiparticle vacuum, like for instance a superfluid or superconducting state
in the BCS approximation \cite{RS.80}, will lead to $S^{\rm sp}>0$, since
$\rho^{\rm sp}$ will be {\it mixed},  i.e., it will have eigenvalues distinct
from $0$ or $1$ (see Appendix). If fermion  quasiparticles are to be allowed,
we can extend the minimization in (\ref{6}) to  {\it all single quasiparticle
basis}, i.e.,
\begin{equation}S^{\rm qsp}=\mathop{\rm Min}_{\bm{a}}S_{\bm{a}}\,,
\label{sqsp1}\end{equation}
where $S_{\bm{a}}=\sum_\nu h(\langle a^\dagger_\nu a_\nu\rangle)$ and $\bm{a}$
denotes a set of fermion operators $a_\nu$ linearly related to the original
operators $c_j$, $c^\dagger_j$ through a general {\it Bogoliubov transformation}
\cite{RS.80}:
\begin{equation} a_\nu=\sum_j \bar{U}_{j \nu} c_j+V_{j\nu}c^\dagger_j\,.
\label{anu}\end{equation}
Eq.\ (\ref{anu}) can be written as
\begin{equation}\left(\begin{array}{c}\bm{a}\\\bm{a}^\dagger\end{array}\right)
={\cal W}^\dagger
\left(\begin{array}{c}\bm{c}\\\bm{c}^\dagger\end{array}\right)\,, \;\;{\cal
W}=\left(\begin{array}{cc}U&V\\\bar{V}&\bar{U}\end{array}\right)\,,
\label{Bg}\end{equation}
where the $2n\times 2n$ matrix ${\cal W}$ should be unitary (i.e.
$UU^\dagger+VV^\dagger=1,\;\;UV^T+VU^T=0$) in order that the operators $a_\nu,
a^\dagger_\mu$ fulfill the fermionic anticommutation relations (1).

One should then consider the {\it extended} $2n\times 2n$ density matrix
\begin{equation}\rho^{\rm qsp}= 1-\langle\left(\begin{array}{c}\bm{c}\\
\bm{c}^\dagger\end{array}\right)\left(\begin{array}{cc}
\bm{c}^\dagger&\bm{c}\end{array}\right)\rangle=
\left(\begin{array}{cc}\rho^{\rm sp}&\kappa\\-\bar{\kappa}\;\;&
1-\bar{\rho}^{\rm sp}\end{array}\right)\label{qsp}\,,\end{equation}
where $\kappa$ is an $n\times n$ antisymmetric matrix containing the pair
annihilation averages
\begin{equation}\kappa_{ij}=\langle c_jc_i\rangle\,,\end{equation}
with  $-\bar{\kappa}_{ij}=\langle c^\dagger_j c^\dagger_i\rangle$ and
$(1-\bar{\rho}^{\rm sp})_{ij}=\langle c_j c^\dagger_i\rangle$. Eq.\ (\ref{qsp})
is a hermitic matrix which can always be diagonalized by a suitable
transformation (\ref{Bg}), such that
\begin{equation}1-\langle\left(\begin{array}{c}\bm{a}\\
\bm{a}^\dagger\end{array}\right)\left(\begin{array}{cc}\bm{a}&\bm{a}^
\dagger\end{array}\right)
\rangle={\cal W}^\dagger\rho^{\rm qsp}{\cal W}= \left(\begin{array}{cc}f&0\\
 0&1-f\end{array}\right)\,,\end{equation}
with $f_{\mu\nu}=f_\nu\delta_{\mu\nu}$ and $f_\nu$, $1-f_\nu$  the eigenvalues
of $\rho^{\rm qsp}$ (which always come in pairs $(f_\nu,1-f_\nu)$, with
$f_\nu\in[0,1]$), entailing
\begin{equation}\langle a^\dagger_\nu a_\mu\rangle=\delta_{\mu\nu} f_\nu\,,
\;\;\;\langle a_\mu a_\nu\rangle=0\,.\end{equation}
It can then be easily shown that the minimum (\ref{sqsp1}) is
\begin{eqnarray}S^{\rm qsp}&=&-\sum_\nu f_\nu\log_2 f_\nu+(1-f_\nu)
\log_2(1-f_\nu)\label{666}\\
 &=&-{\rm tr}'\,\rho^{\rm qsp}\log_2 \rho^{\rm qsp}\,.\label{sqsp2}
\end{eqnarray}
where ${\rm tr}'$ denotes the trace in the extended sp space. \\
{\it Proof:}
Since both $p_j=\langle c^\dagger_j c_j\rangle$ and $1-p_j$ are the diagonal
elements of $\rho^{\rm qsp}$, denoting with $q_j$ and $\lambda_\nu$ the full
set of diagonal elements and eigenvalues of $\rho^{\rm qsp}$, we obtain
$q_j=\sum_\nu |{\cal W}_{j\nu}^2|\lambda_\nu$ and hence, due to the concavity
of $f(p)=-p\log_2 p$, $S_{\bm c}=\sum_j f(q_j)\geq \sum_{j,\nu}|{\cal
W}_{j\nu}^2| f(\lambda_\nu)=\sum_\nu f(\lambda_\nu)=S^{\rm qsp}$. \qed.

Eq.\ (\ref{sqsp2}) vanishes iff $f_\nu$ is either 0 or $1$ for all $\nu$,
i.e., {\it iff $|\Psi\rangle$ is a particle or quasiparticle Slater determinant}. By an
elementary particle-hole transformation we can always change such state to a
quasiparticle vacuum, so that we can say $S^{\rm qsp}=0$ iff $|\Psi\rangle$ is
a quasiparticle vacuum. In other words, $S^{\rm qsp}=0$ iff there is a single
quasiparticle basis where {\it every} level is disentangled from the rest.

Eq.\ (\ref{sqsp1}) also  measures the mixedness of $|\Psi\rangle$ with respect
to the set of all {\it generalized  one-body operators}, of the form
\begin{eqnarray}
O&=&\sum_{i,j} o^{11}_{ij}c^\dagger_i c_j +{\textstyle\frac{1}{2}}(
o^{20}_{ij}c_i c_j+o^{02}_{ij}c^\dagger_i c^\dagger_j)
-{\textstyle\frac{1}{2}}{\rm tr}\,o^{11}\label{Og1}\\
&=&{\textstyle\frac{1}{2}\left(\begin{array}{cc}\bm{c}^\dagger&\bm{c}\end{array}\right)
{\cal O}\left(\begin{array}{c}\bm{c}\\\bm{c}^\dagger\end{array}\right)\,,\;\;
{\cal O}=\left(\begin{array}{cc}o^{11}&o^{02}\\o^{20}&-(o^{11})^T\end{array}\right)}\,,
 \end{eqnarray}
i.e.,  general quadratic functions of $\bm{c}$, $\bm{c}^\dagger$ (the  constant
term in (\ref{Og1}) is just added for convenience), since their averages are
completely determined by $\rho^{\rm qsp}$:
\begin{equation}
\langle\Psi|O|\Psi\rangle=
{\rm tr}[\rho^{\rm sp}\,o^{11}-{\textstyle\frac{1}{2}}o^{11}+
{\textstyle\frac{1}{2}}(\kappa o^{20}
-\bar{\kappa}o^{02})]={\textstyle\frac{1}{2}}{\rm tr}'\,\rho^{\rm qsp}{\cal O}\,.
\end{equation}
The present scheme allows then to properly treat states which do not have a
definite fermion number and lead to non-zero contractions $\langle c_i
c_j\rangle$. The whole  formalism becomes then strictly invariant under
arbitrary  particle hole transformations (\ref{3}) applied  to some subset
of levels, which will move elements from $\rho^{\rm sp}$ to $\kappa$ and
viceversa, but which will not alter the spectrum of $\rho^{\rm qsp}$. The
latter remains actually invariant under {\it arbitrary quasiparticle unitary
transformations} $|\Psi\rangle\rightarrow \exp[-iO]|\Psi\rangle$, where $O$ is
an hermitian generalized one-body operator of the form (\ref{Og1}), since they
just lead to a unitary transformation of $\rho^{\rm qsp}$, i.e., $\rho^{\rm
qsp}\rightarrow {\cal W}\rho^{\rm qsp}{\cal W}^\dagger$, with ${\cal
W}=e^{-i{\cal O}}$.

We notice that a transformation $a_\nu\leftrightarrow a^\dagger_\nu$ obviously
changes $f_\nu\leftrightarrow 1-f_\nu$, so that there is no unique way to
select which of the eigenvalues of $\rho^{\rm qsp}$ will be the $f_\nu$'s or
the $1-f_\nu's$.  One can choose the $f_\nu$'s as the lowest eigenvalues (such
that $|\Psi\rangle$ becomes a quasiparticle vacuum when $S^{\rm qsp}=0$), but
it is also possible to set  ${\rm Det}[U]\neq 0$, which ensures that the vacuum of the
$a_\nu$  has the same number parity as $|0\rangle$ (Eq.\ (\ref{0a})).
These choices do not affect the entropy $S^{\rm qsp}$. We also remark that the
maximally entangled state, i.e., that with maximum $S^{\rm qsp}$, corresponds
to the exceptional case $f_\nu=1/2$ $\forall$ $\nu$, where $S^{\rm qsp}=n$ and
$\rho^{\rm qsp}=I_{2n}/2$ becomes proportional to the identity matrix, remaining
then invariant under {\it any} Bogoliubov transformation.

\subsection{Generalized entropic inequalities and quadratic entropy}
From their definitions, it follows that the  entropies (\ref{5}),
(\ref{66}) and (\ref{sqsp2}) satisfy the inequality chain
\begin{equation}S_{\bm c}\geq S^{\rm sp}\geq S^{\rm qsp}\,.\label{ineq1}\end{equation}
Eq.\ (\ref{ineq1}) actually holds for more general entropic forms. If
$\tilde{\rho}^{\rm sp}=\left(\begin{array}{cc}\rho^{\rm sp}&0\\0&1-\rho^{\rm
sp}\end{array}\right)$ is the  extended $\rho^{\rm sp}$ and  $\tilde{\rho}^{\rm
sp}_d$ the diagonal of $\tilde{\rho}^{\rm sp}$, we obtain, with the same
previous arguments,
\begin{equation}
S_f(\tilde{\rho}_d^{\rm sp})\geq S_f(\tilde{\rho}^{\rm sp})\geq S_f(\rho^{\rm qsp})\,,
\label{Sf2}\end{equation}
where
\begin{equation}S_f(\rho)={\rm tr}\,f(\rho)\,,\label{Sf}\end{equation}
 with  $f:[0,1]\rightarrow\mathbb{R}$ a strictly concave function satisfying
 $f(0)=f(1)=0$, represents a generalized entropic form \cite{CR.02,GR.14}.
Moreover, these matrices fulfill the majorization relation \cite{Bh.97}
\begin{equation}\tilde{\rho}_d^{\rm sp}\prec\tilde{\rho}^{\rm sp}\prec\rho^{\rm qsp}
 \label{prec2}\,,\end{equation}
where $\rho\prec\rho'$ means  here $\sum_{i=1}^j \lambda_i \leq \sum_{i=1}^j
\lambda'_i$ for $j=1,\ldots, 2n-1$,  with $\lambda_i$, $\lambda'_i$ the
eigenvalues of $\rho$ and $\rho'$ sorted in decreasing order, since the sorted
set of diagonal elements in an orthonormal basis of an hermitian operator are
always majorized by the sorted set of its eigenvalues \cite{Bh.97}. Eq.\
(\ref{prec2}) allows to extend (\ref{Sf2}) to any Schur-concave function
\cite{Bh.97} of the extended density matrices.

A particularly useful example, which will play an important role in the next section,
is the quadratic entropy $S_2(\rho)$ (also denoted as linear entropy),
obtained for $f(p)=2p(1-p)$:
\begin{eqnarray}
S_2(\rho^{\rm qsp}) &=&
2\,{\rm tr}'\,[\rho^{\rm qsp}(1-\rho^{\rm qsp})]\nonumber\\
&=&4\,{\rm tr}[\rho^{\rm sp}(1-\rho^{\rm sp})-\kappa^\dagger\kappa]\label{C2}\\
 &=&4\sum_{\nu} f_\nu (1-f_\nu)\label{C3}\,,\end{eqnarray}
where the factor $2$ has been chosen such that its maximum value for a
single level is $1$. Unlike the von Neumman entropy (\ref{sqsp2}),
$S_2(\rho^{\rm qsp})$ can be
evaluated just by taking the trace in (\ref{C2}), without explicit knowledge of
the eigenvalues $f_\nu$ of $\rho^{\rm qsp}$. Yet, like $S^{\rm qsp}$, it is
non-negative and vanishes iff $|\Psi\rangle$ is a quasiparticle vacuum or
Slater Determinant. Eq.\ (\ref{Sf2}) implies in particular $\sum_j
p_j(1-p_j)\geq \sum_k p'_k(1-p'_k)\geq \sum_\nu f_\nu(1-f_\nu)$.

\subsection{Mixed states}
Let us now consider mixed fermion states, assumed as convex mixtures of pure
states with definite  number parity, i.e.,
\begin{equation} \rho=\sum_i q_i|\Psi_i\rangle\langle \Psi_i|\,,\end{equation}
where  $q_i\geq 0$, $\sum_i q_i=1$ and $P|\Psi_i\rangle=\pm|\Psi_i\rangle$,
such that $[\rho,P]=0$. We can define an entanglement measure for these mixed
states in a way similar to the entanglement of formation \cite{WW.89,BD.96},
through the convex roof extension of $S^{\rm qsp}$,
\begin{equation}
E^{\rm qsp}(\rho)=\mathop{\rm Min}_{\{q'_i,|\Psi'_i\rangle\}}
 \sum_i q'_i S^{\rm qsp}(|\Psi'_i\rangle)\,,\label{Eqsp}
\end{equation}
where $\rho=\sum_i q'_i|\Psi'_i\rangle\langle \Psi'_i|$, $q'_i\geq 0$, and  the
minimization is over all  decompositions of $\rho$ as  convex mixtures of pure
states, assumed again of definite number parity. Eq.\ (\ref{Eqsp}) vanishes {\it iff
$\rho$ is a convex mixture of particle or quasiparticle Slater determinants},
i.e., of suitable quasiparticle vacua, and reduces to $S^{\rm qsp}$ for pure states.
This quantity will be evaluated exactly in the particular system of the next section.

As a general application of $E^{\rm qsp}$,  let us consider an interacting
fermion system at finite temperature $T$. For attractive two-body couplings,
the static path approximation \cite{SPA0,SPA}
will lead to a classically correlated density operator $\rho_{\rm
SPA}$, which is a convex mixture of (non-commuting) thermal states diagonal in a
basis of particle or quasiparticle Slater determinants, associated with
different values of the running effective order parameters. Hence, $E^{\rm
qsp}(\rho_{\rm SPA})=0$, in agreement with the fact that $\rho_{\rm SPA}$
contains just static fluctuations around mean field. Such correlated but still
unentangled approximation can be derived from the auxiliary field path
integral representation \cite{HS}, and becomes exact at
sufficiently high $T$ \cite{SPA}. Its breakdown at low $T$
reflects the onset of entanglement, i.e., of a finite value of $E^{\rm
qsp}(\rho)$. Eq.\ (\ref{Eqsp}) defines a limit temperature $T_L$ above which
$E^{\rm qsp}=0$.  Mixtures of fermionc gaussian states are also important in
noisy fermionic quantum computation models \cite{Os.14,FM.13}.

 \section{The case of four single particle levels}
We will now examine in detail the special case  of a fermion system with single
particle space dimension $n=4$.  This is the  lowest dimension where
non-trivial fermionic entanglement arises, i.e., where $S^{\rm qsp}$ can be
non-zero,  as will be verified. We will extend the results of \cite{Sch.01},
which considered just pure or mixed states with a definite fermion number, to
general states which do not necessarily have a definite fermion number, yet
still have a definite number parity $P$ (see also \cite{Os.14,SL.14}). This sp
space can accommodate 8 linearly independent pure states of the same number
parity, so that the Hilbert space dimension for fixed $P$ is 8.

\subsection{Pure states }
\subsubsection{Odd parity states}
\begin{figure}[t]
\vspace*{0cm}

\centerline{\hspace*{-0.2cm}\scalebox{.4}{\includegraphics{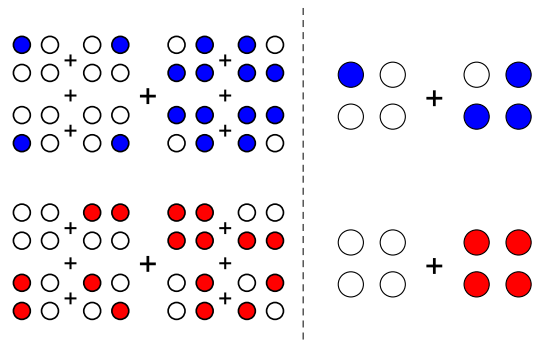}}}
\caption{(Color online) Schematic representation of pure fermion states  with
odd (top) or even (bottom)  number parity. A general state with definite number
parity is a superposition of the 8 states indicated on the left of the dashed
vertical line, where a full disk indicates an occupied level. In the normal
form ($\rho^{\rm qsp}$ diagonal), obtained after a suitable Bogoliubov
transformation,  it can be reduced to the superposition of two states  like
those indicated on the right. The state is entangled (in the sense of not being
a quasiparticle vacuum or Slater Determinant) iff the product $C$ (Eq.\
(\ref{eta})) of the coefficients of the left and right groups of four states is
non-zero, implying nonzero weight for both states of the normal
representation.} \label{f1}
\end{figure}

We first consider pure states $|\Psi\rangle$ of this system with odd number
parity: $P|\Psi\rangle=-|\Psi\rangle$. These states are then linear
combinations of single fermion states and three-fermion states, so a general
odd state can be written as (Fig.\ \ref{f1}, top)

\begin{equation}
|\Psi\rangle=\sum_{i=1}^4 (\alpha_i c^\dagger_i|0\rangle+\bar{\beta}_i c_i|\bar 0\rangle)\,,
\label{odd}
\end{equation}
where  $|\bar{0}\rangle=c^{\dagger}_{1}c^{\dagger}_{2}c^{\dagger}_{3}c^{\dagger}_{4}|0\rangle$
is the completely occupied state and $|\bm{\alpha}|^2+|\bm{\beta}|^2=1$, with
$\bm{\alpha}$, $\bm{\beta}$ four dimensional complex vectors.  It is easily
seen that the single hole states $c_i|\bar{0}\rangle$ are
\begin{equation}c_i |\bar{0}\rangle=\frac{1}{3!}\sum_{j,k,l} \epsilon_{ijkl}c^{\dagger}_{j}
c^{\dagger}_{k}c^{\dagger}_{l}|0\rangle\,,\end{equation}
where $\epsilon_{ijkl}$ denotes the completely antisymmetric Levi-Civita tensor
in dimension $4$. The elements of the generalized one-body density matrix (\ref{qsp})
are then given by
\begin{eqnarray} \rho^{\rm sp}_{ij}&=&\langle c^\dagger_j c_i\rangle
=\alpha_i\bar{\alpha}_j-{\beta}_i \bar{\beta}_j+|\bm{\beta}|^2\delta_{ij}\,,
\label{rhos1}\\ \kappa_{ij}&=&\langle c_j c_i\rangle=\sum_{k,l}
\epsilon_{ijkl}\bar{\alpha}_l \bar{\beta}_k\,,\label{rhos2}
\end{eqnarray}
i.e.,  $\rho^{\rm
sp}=\bm{\alpha}\bm{\alpha}^\dagger-\bm{\beta}\bm{\beta}^{\dagger}+|\bm{\beta}|^2
I_4$. We now show that the ensuing eigenvalues $f_\nu$ of the $8\times 8$
matrix $\rho^{\rm qsp}$ are {\it four-fold degenerate} and given by
\begin{equation}
f_\pm=\frac{1\pm\sqrt{1-C^2(|\Psi\rangle)}}{2}\label{la}\,,
\end{equation}
 where $C(|\Psi\rangle)$ is fully determined by the $S_2$ entropy (\ref{C2}),
\begin{eqnarray}
C(|\Psi\rangle)&=&\sqrt{S_2(\rho^{\rm qsp})/4}
=\sqrt{{\rm tr}\,[\rho^{\rm sp}
(I_4-\rho^{\rm sp})-\kappa^\dagger\kappa]}\nonumber\\
&=&2|{\bm{\beta}}^\dagger\bm{\alpha}|
 =2|\sum_{i=1}^4 \bar{\beta}_i\alpha_i|
\label{eta}\,,
\end{eqnarray}
and plays the role of a pure state fermionic concurrence. It satisfies
$0\leq C\leq 1$, and  as will be seen in the next subsection, it is the
generalization of the Slater correlation measure defined in
\cite{Sch.01,Eck.02} for two fermion states. It also coincides with the
quadratic invariant derived in \cite{SL.14} using a spinors classification
based approach. The entanglement entropy (\ref{666}) becomes
\begin{equation}
S^{\rm qsp}=4h(f_+)=-4(f_+\log_2 f_+ +f_ -\log_2 f-)\,.
\label{Sqsp2}\end{equation}
{\it Proof.} We first consider a unitary transformation $\bm{c}\rightarrow U
\bm{c}$ of the operators $c_j$, such that \begin{equation}
\bm{\alpha}\rightarrow U^\dagger \bm{\alpha},\;\;\bm{\beta}\rightarrow {\rm
 Det}[U^\dagger]  U^\dagger \bm{\beta}\,,\label{U1}\end{equation}
in (\ref{odd}), which does not affect the value of $C(|\Psi\rangle)$ (Eq.\
(\ref{eta})). By choosing an orthonormal basis of $\mathbb{C}^4$ such that the
original vectors $\bm{\alpha}$ and $\bm{\beta}$ are generated by the first two
elements (for instance, $\bm{e}_1\propto \bm{\alpha}$ and $\bm{e}_2\propto
\bm{\beta}-(\bm{\alpha}^\dagger\bm{\beta}) \bm{\alpha}/|\bm{\alpha}|^2$), we
can use this first transformation to set $\alpha_3=\alpha_4=0$,
$\beta_3=\beta_4=0$ in the new basis. In this case, Eqs.\
(\ref{rhos1})--(\ref{rhos2}) lead to
\begin{eqnarray}
\rho^{\rm sp}&=&\begin{pmatrix} |\alpha_1|^2+|\beta_2|^2&\alpha_1\bar{\alpha}_2
-\beta_1\bar{\beta}_2&0&0
\\\alpha_2\bar{\alpha}_1-\beta_2\bar{\beta}_1&|\alpha_2|^2
+|\beta_1|^2&0&0\\0&0&|\bm{\beta}|^2&0\\0&0&0&|\bm{\beta}|^2\end{pmatrix},\label{rh1}\\
\kappa&=&\begin{pmatrix} 0&0&0&0\\ 0&0&0&0\\ 0& 0&0& \bar{\alpha}_2\bar{\beta}_1
-\bar{\alpha}_1\bar{\beta}_2
\\0&0&\bar{\alpha}_1\bar{\beta}_2-\bar{\alpha}_2\bar{\beta}_1&0\end{pmatrix}\,.
\label{rh2}\end{eqnarray}
It is then seen that the diagonalization of $\rho^{\rm qsp}$ is achieved
through i) a unitary transformation of the operators $c_1$, $c_2$,
\begin{equation} c_1=ua_1+va_2\,,\;\;c_2=-\bar{v}a_1 +ua_2\,,\end{equation}
with $^{\;u}_{|v|}=\sqrt{\frac{f_+-f_-\pm 2\epsilon}{2(f_+-f_-)}}$ and
$\epsilon=|\alpha_1^2|+|\beta_2^2|-\frac{1}{2}$, which diagonalizes the first
$2\times 2$ block of $\rho^{\rm sp}$ and $1-\bar{\rho}^{\rm sp}$,  plus ii) a
Bogoliubov transformation of the operators $c_3$, $c_4$,
\begin{equation} c_3=u' a_3+v' a^\dagger_4\,,\;\;c^\dagger_4=-\bar{v}'a_3+u' a^\dagger_4\,,
\label{uv2}\end{equation}
with $^{\;u'}_{|v'|}=\sqrt{\frac{f_+-f_-\pm 2\epsilon'}{2(f_+-f_-)}}$and
$\epsilon'=|\bm \beta^2|-\frac{1}{2}$, which diagonalizes the rest of
$\rho^{\rm qsp}$, comprising again two $2\times 2$ blocks
$(^{|\bm{\beta}|^2\;\pm\kappa_{34}}_{\pm\bar{\kappa}_{34}\;|\bm{\alpha}|^2})$.
These four $2\times 2$ blocks have all trace $1$ and determinant
$C^2(|\Psi\rangle)/4$, leading then to the {\it same} eigenvalues $f_{\pm}$ of
Eq.\ (\ref{la}) (a $2\times 2$ matrix with trace $t$ and determinant $d$ has
eigenvalues $\frac{t\pm\sqrt{t^2-4d}}{2}$). \qed

Note from (\ref{rh2}) that if $\rho^{\rm qsp}$ is diagonal ($\rho^{\rm sp}$
diagonal and $\kappa=0$) and $C(|\Psi\rangle)<1$, then necessarily
$\alpha_2=\beta_2=0$ or $\alpha_1=\beta_1=0$ in (\ref{rh2}).  This implies that
after the previous transformations,  $|\Psi\rangle$ can be written in the
normal form (top right scheme in Fig.\ \ref{f1})
\begin{equation}
|\Psi\rangle= \alpha' a^\dagger_1|0_{\bm{a}}\rangle+
\bar{\beta}'a_1|\bar{0}_{\bm{a}}\rangle\label{nl}\,,
\end{equation}
i.e., $\bm{\beta}'\propto\bm{\alpha}'$, with $|0_{\bm{a}}\rangle$ the vacuum of
the $\bm{a}$ operators, $|\bar{0}_{\bm{a}}\rangle=a^\dagger_1 a^\dagger_2
a^\dagger_3 a^\dagger_4|0_{\bm{a}}\rangle$, and $|\alpha'|^2=f_+$,
$|\beta'|^2=f_-$ if $|\alpha'|\geq |\beta'|$, such that
$C(|\Psi\rangle)=2|\alpha'\bar{\beta}'|$. This state leads to
  \[\rho^{\rm qsp}_a=1-\langle \begin{pmatrix}\bm{a}\\
  \bm{a}^\dagger\end{pmatrix}\begin{pmatrix}\bm{a}^\dagger &\bm{a}\end{pmatrix}\rangle=
    \begin{pmatrix}|\alpha'|^2&0&0&0\\0&|\beta'|^2I_3&0&0\\0&0&|\beta'|^2&0\\
 &0&0&|\alpha'|^2I_3\end{pmatrix}\,.\]
On the other hand, in the maximally entangled case $C(|\Psi\rangle)=1$,
$f_{\pm}=1/2$ and $\rho^{\rm qsp}=I_8/2$ in {\it any} basis, i.e., after {\it
any} Bogoliubov transformation. In this case $\bm{\beta}=e^{i\phi}\bm{\alpha}$,
with $|\bm{\alpha}|=|\bm{\beta}|=1/\sqrt{2}$, and the form (\ref{nl}) is
obtained just by choosing $\bm{e}_1$ in the direction of $\bm{\alpha}$.

It is apparent that if $\bm{\beta}=\bm{0}$ in (\ref{odd}),  $|\Psi\rangle$ can
be written a single fermion state $a^\dagger_1|0\rangle$, where
$a^\dagger_1=\sum_i \alpha_i c^\dagger_i$. Similarly, if $\bm{\alpha}=\bm{0}$
$|\Psi\rangle$ can be written a single hole state $a_1|\bar{0}\rangle$, with
$a_1=\sum_i \bar{\beta}_i c_i$. Accordingly, $C(|\Psi\rangle)=0$ in these
cases. The vanishing of $C(|\Psi\rangle)$ for  nonzero but {\it orthogonal}
$\bm{\alpha}$ and $\bm{\beta}$ (Eq.\ (\ref{eta})) generalizes the previous
result, showing that in this case $|\Psi\rangle$ can still be written as single
quasiparticle ($\beta'=0$) or  quasihole ($\alpha'=0$) after a suitable
Bogoliubov transformation of the original operators.  This includes the three
level case where, for instance, the fourth level is empty, which implies
$\alpha_4=0$ and $\beta_i=0$ for $i=1,2,3$, leading necessarily to
$\bm{\beta}^\dagger\bm{\alpha}=0$.

We also mention that the four eigenvalues of $\rho^{\rm sp}$ in Eq.\ (\ref{rh1})  are
$f_{\pm}$ and $|\bm{\beta}|^2$, the latter two-fold degenerate.
Since  $C(|\Psi\rangle)\leq 2|\bm{\alpha}| |\bm{\beta}|$,
\[f_+\geq
\frac{1+\sqrt{1-4|\bm{\alpha}|^2|\bm{\beta}|^2}}{2}={\rm Max}
 [|\bm{\alpha}|^2,|\bm{\beta}|^2]\,,\]
being then verified that the eigenvalues of $\rho^{\rm sp}$ are majorized by those
of $\rho^{\rm qsp}$ and hence, that $S^{\rm sp}\geq S^{\rm qsp}$, $S_2^{\rm
sp}\geq S_2^{\rm qsp}=4C^2(|\Psi\rangle)$.

{\it Dualization.} Eqs.\ (\ref{odd}) and (\ref{eta}) indicate that the state
$c_i|\bar{0}\rangle$ plays the role of partner or dual of the state
$c^\dagger_i|0\rangle$. We may obtain the partner state with the
hermitian operator
\begin{equation}T=-\frac{1}{3!}\sum_{i,j,k,l}
\varepsilon_{ijkl}[c^\dagger_i c^\dagger_j c^\dagger_k c_l +c^\dagger_i c_j c_k c_l]
\label{Todd}\end{equation}
such that for $i=1,\ldots,4$, $T c^\dagger_i|0\rangle=c_i|\bar{0}\rangle$,
$Tc_i|\bar{0}\rangle=c^\dagger_i|0\rangle$.
We can then express Eq.\ (\ref{eta}) as
\begin{equation} C(|\Psi\rangle)=|\langle \tilde{\Psi}|\Psi\rangle|,
\;\;|\tilde{\Psi}\rangle=T|\Psi\rangle^*\label{eta2}\end{equation}
where $|\Psi\rangle^*=\sum_j
\bar{\alpha}_ic^\dagger_i|0\rangle+\beta_ic_i|\bar{0}\rangle$ denotes the
``conjugated'' state in this basis. Note that the $8\times 8$ matrix that
represents $T$ in the basis ($c^\dagger_1|0\rangle,\ldots,c^\dagger_4|0\rangle,
c_1|\bar{0}\rangle,\ldots,c_4|\bar{0}\rangle)$
 is just
\begin{equation}T=\begin{pmatrix}0&I_4\\I_4&0\end{pmatrix}\,.\label{T}\end{equation}
A generalization of (\ref{Todd}) for higher dimensions is considered in \cite{SL.14}.
\subsubsection{Even parity states}
We now consider pure states  of even number parity,
$P|\Psi\rangle=|\Psi\rangle$. They can be obtained, for instance, by changing
a particle for a hole in the odd-parity states. An even state is then a linear
combination of the eight states shown in the bottom plots  of Fig.\ \ref{f1},
comprising the vacuum $|0\rangle$, six two-fermion states  and the completely
full state
$|\bar{0}\rangle=c^\dagger_1c^\dagger_2c^\dagger_3c^\dagger_4|0\rangle$.  We
can write this state as
\begin{equation}|\Psi\rangle=\alpha_1|0\rangle-\bar{\beta}_1|\bar{0}\rangle+\sum_{j=2}^4
\alpha_j c^\dagger_jc^\dagger_1|0\rangle+\bar{\beta}_j c_1
 c_j|\bar{0}\rangle\,, \label{even}\end{equation}
which is just Eq.\ (\ref{odd}) with the replacements
$c^\dagger_1\leftrightarrow c_1$ and $|0\rangle\leftrightarrow
c^\dagger_1|0\rangle$, implying
$|\bar{0}\rangle\leftrightarrow-c_1|\bar{0}\rangle$. Notice that
\begin{equation}
c_1 c_j|\bar{0}\rangle=\frac{1}{2!}\sum_{k,l}
\epsilon_{j1kl}c^\dagger_k c^\dagger_l |0\rangle\,.
\end{equation}
In this notation, the eigenvalues of $\rho^{\rm qsp}$ are then given by Eq.\
(\ref{la}) with the same expression (\ref{eta}) for $C(|\Psi\rangle)$. The
entanglement entropy $S^{\rm qsp}$ is given again by Eq.\ (\ref{Sqsp2}).
Notice, however, the minus sign in the term associated with $\bar{\beta}_1$.
The expression (\ref{eta}) reduces to that of \cite{Sch.01} for the case of
two-fermion states ($\alpha_1=\beta_1=0$).

The state (\ref{even}) is then a Slater Determinant or quasiparticle vacuum iff
$C(|\Psi\rangle)=0$. As a check, the quasiparticle vacuum (\ref{0a})
corresponds in the present case to
\begin{eqnarray}\bm{\alpha}&\propto& (1,T_{21}, T_{31}, T_{41}),\nonumber\\
 \bar{\bm{\beta}}&\propto&(-T_{21}T_{43}-T_{31}T_{24}-T_{41}T_{32},T_{43},T_{24},T_{32})
\,,\end{eqnarray}
being verified that $\sum_{i=1}^4 \bar{\beta}_i \alpha_i=0$. It is also
seen that in the three-level case (i.e., level $4$ empty, implying
$\alpha_4=0$ and $\beta_j=0$ for $j=1,2,3$) $C(|\Psi\rangle)$ is always zero.

The normal form (\ref{nl}) becomes here
\begin{equation}|\Psi\rangle=\alpha'|0_{\bm{a}}\rangle-\bar{\beta}'|\bar{0}_{\bm{a}}
 \rangle\label{nl2}\,,\end{equation}
i.e., {\it a superposition of the vacuum and the maximally occupied state}
(bottom right scheme in Fig.\ \ref{f1}) for the diagonalizing quasiparticle
operators. Of course, after a trivial particle hole transformation
$a_j\leftrightarrow a^\dagger_j$ for $j=1,2$, we may always rewrite (\ref{nl2})
as a sum of two two-fermion states, i.e.,
\begin{equation}|\Psi\rangle=\alpha'a^\dagger_2a^\dagger_1|0_{\bm{a}}\rangle+
\bar{\beta}' a^\dagger_4 a^\dagger_3|0_{\bm{a}}\rangle\label{nl22}\,,\end{equation}
which extends the results of \cite{Sch.01} valid for two-fermion states to
arbitrary definite parity states.

The dualization operator (\ref{Todd}) takes here the form
\begin{equation}T=-c^\dagger_1 c^\dagger_2 c^\dagger_3 c^\dagger_4 -c_4 c_2 c_3 c_1-
\frac{1}{4} \sum_{i,j,k,l}\epsilon_{ijkl}c^\dagger_i c^\dagger_j c_k c_l\,,
\label{Teven}\end{equation}
which satisfies
\[T|0\rangle=-|\bar{0}\rangle,\;T|\bar{0}\rangle=-|0\rangle,\;
Tc^\dagger_ic^\dagger_j|0\rangle=\frac{1}{2}
\sum_{k,l}\epsilon_{ijkl}c^\dagger_kc^\dagger_l|0\rangle\,,\] i.e.,
$Tc^\dagger_ic^\dagger_1|0\rangle=c_1c_i|\bar{0}\rangle$,
$Tc_1c_i|\bar{0}\rangle=c^\dagger_ic^\dagger_1|0\rangle$. It is represented in
the special basis $\{|0\rangle, c^\dagger_2 c^\dagger_1|0\rangle,$ $c^\dagger_3
c^\dagger_1|0\rangle, c^\dagger_4 c^\dagger_1|0\rangle, -|\bar{0}\rangle,
c^\dagger_4 c^\dagger_3|0\rangle,c^\dagger_2 c^\dagger_4|0\rangle, c^\dagger_3
c^\dagger_2|0\rangle\}$ by {\it the same matrix} (\ref{T}). We can then write
again $C(|\Psi\rangle)$  in the form  (\ref{eta2}). If $\alpha_1=\beta_1=0$,
the ensuing expression reduces to that of \cite{Sch.01}.

The two-fermion states considered in \cite{Sch.01,Eck.02} are only a particular
case of the more general even states (\ref{even}). For two fermion states the
contractions $\langle c_ic_j\rangle$ obviously vanish ($\kappa=0$), and the
eigenvalues $f_\nu$ of the generalized one body density matrix $\rho^{\rm qsp}$
reduce to those of the one body density matrix $\rho^{\rm sp}$, implying
$S^{\rm sp}=S^{\rm qsp}$.

\subsection{Mixed states and analytic evaluation of the concurrence}
The fermionic concurrence for mixed states can be defined by the convex roof
extension of Eq.\ (\ref{eta}). For two-fermion states an explicit expression
was derived in \cite{Sch.01}. We will here extend this expression to the
present general states (see also \cite{Os.14}). Let
\begin{equation}\rho=\sum_k \lambda_k|\Psi_k\rangle\langle \Psi_k|
\end{equation}
be a mixed state with eigenvectors $|\Psi_k\rangle$
and eigenvalues $\lambda_k$, with  $\lambda_k>0$ for $k=1\ldots,r$ and $r\leq
8$ the rank of $\rho$.  We will assume that all $|\Psi_k\rangle$ have the
same number parity,
such that they are of the form (\ref{odd}) or (\ref{even}), i.e.,
$|\Psi_k\rangle=\sum_{i=1}^4 \alpha_{ki}
c^\dagger_i|0\rangle+\bar{\beta}_{ki}c_i|\bar{0}\rangle$ in the odd case.
Every convex decomposition $\rho=\sum_{j=1}^{r'} p_j|\Phi_j\rangle\langle\Phi_
j|$ can be obtained
from these eigenvectors through a $r'\times r$ matrix $U$ with orthonormal
columns ($U^\dagger U=I_r$) such that
$\sqrt{p_j}|\Phi_j\rangle=\sum_{k=1}^rU_{jk}\sqrt{\lambda_k}|\Psi_k\rangle$.
Note that the states $|\Phi_j\rangle$ are normalized, so that
$p_j=\sum_{k=1}^r \lambda_k |U_{jk}|^2$.

The average fermionic concurrence (generalized Slater measure) of such
decomposition is
\begin{eqnarray}
\langle C(\{p_j,|\Phi_j\rangle\})\rangle&=&\sum_j p_j C(|\Phi_j\rangle)
=\sum_j p_j|\langle\tilde\Phi_j|\Phi_j\rangle|\nonumber\\
&=&\sum_j |\sum_{k,l}U_{jk}
U_{jl}\sqrt{\lambda_k\lambda_l}\langle\tilde \Psi_k|\Psi_l\rangle|\,.
\label{Avslat}
\end{eqnarray}
The matrix $C$ of elements
\begin{equation}C_{kl}=\sqrt{\lambda_k\lambda_l}\langle\tilde \Psi_k|\Psi_l
\rangle=
 \sqrt{\lambda_k\lambda_l}(\bm{\beta}_l^\dagger\bm\alpha_k+
\bm{\beta}^\dagger_k\bm \alpha_l)\,,
 \end{equation}
is complex symmetric. Therefore, it admits a decomposition of the form \cite{Sch.01}
$C=VDV^T$, where $V$ is a unitary matrix and $D$ is a real diagonal matrix
whose diagonal elements $d_k\geq 0$ are the square root of the eigenvalues of
$CC^\dagger=C\bar{C}$, sorted in {\it descending} order. Defining $S=UV$, Eq.\
(\ref{Avslat}) then reads
\begin{equation}
\langle C(\{p_j,|\Phi_j\rangle\})\rangle=\sum_j|\sum_{k}S^2_{jk}d_k|\,.
\label{av2}
\end{equation}
Since $\sum_j|\sum_{k}S^2_{jk}d_k|\geq \sum_j(d_1|S^2_{j1}|
-\sum_{k\geq 2}|S^2_{jk}|d_k)=d_1-\sum_{k\geq 2} d_k$,
a necessary condition for the ``separability'' of $\rho$, i.e., for $\rho$ to
be a convex mixture of Slater determinants with the same number parity,  is
\begin{equation}
d_1\leq\sum_{k\geq 2}d_k.\label{Sepcond}
\end{equation}
As in the case of two fermion states, we will now show, following the scheme of
\cite{Sch.01}, that this is also a sufficient condition for separability.
Indeed, from (\ref{av2}) it is seen that $\rho$ is separable
if there is a matrix $S$ with orthonormal columns such that for every $j$,
\begin{equation}
|\sum_{k=1}^r d_k S_{jk}^2|=0.
\end{equation}
Now, provided condition (\ref{Sepcond}) is fulfilled, there are always phases
${\theta_k}, k=2,..,r$ such that $d_1=|\sum_{k=2}^r d_k e^{i\theta_k}|$. Then a
matrix with elements $S_{jk}=\frac{e^{i(\theta_k+\mu_{jk}\pi)}}{\sqrt{r'}}$,
where $\mu_{jk}=0,1$ and $\theta_1=0$, will give the desired decomposition if
the signs $e^{i\mu_{jk}\pi}$ can be arranged such that the condition $S^\dagger
S=I_{r}$ is satisfied. This can be ensured by
taking $r'=2$ if $r=2$, $r'=4$ if $r=3,4$ \cite{Sch.01} and $r'=8$ if $5\leq r
\leq 8$, where
we can set $\mu_{j1}=0$ $\forall$ $j$ and $(\mu_{1k},\ldots,\mu_{8k})$ as
$(0,0,0,0,1,1,1,1)$,$(0,0,1,1,0,0,1,1)$,
$(0,0,1,1,1,1,0,0)$,$(0,1,0,1,0,1,0,1)$,$(0,1,0,1,1,0,1,0)$,
$(0,1,1,0,0,1,1,0)$,$(0,1,1,0,1,0,0,1)$ for $k=2,\ldots,8$.
This completes the proof.

On the other hand, if condition (\ref{Sepcond}) does not hold, the average
(\ref{av2}) is not smaller than $d_1-\sum_{k=2}^r d_k$. This lower bound may be
achieved with the same construction used above, choosing $\theta_k=\pi/2$ for
$k\geq 2$. Then,
the minimizing decomposition is that where all the components have the same
concurrence, which is the concurrence of the state $\rho$,
\begin{equation}
C(\rho)={\rm Min}_{\{p_j,|\Phi_j\rangle\}}\sum_j p_j
C(|\Phi_j\rangle)={\rm Max}[d_1-\sum_{k=2}^r d_k,0]\,.\label{Cr}
\end{equation}
Using the dualization matrix (\ref{T}) we may also obtain the eigenvalues
$d_k$ as those of
\begin{equation}R=\sqrt{\rho^{1/2} T\rho^*T\rho^{1/2}}\,,\end{equation}
where $\rho^*$ means conjugation in the basis where $T$ takes the form
(\ref{T}).

Once $C$ is obtained, we can evaluate the convex roof extension  (\ref{Eqsp})
of $S^{\rm qsp}$ as
\begin{equation}E^{\rm qsp}(\rho)=4h{\textstyle\left(\frac{1+\sqrt{1-C^2(\rho)}}{2}
 \right)}\,,\label{Er}\end{equation}
in the same way as in the two-qubit case \cite{W.98}, since for pure states we
have similarly $S^{\rm qsp}=4h(\frac{1+\sqrt{1-C^2(|\Psi\rangle)}}{2})$ (Eq.\
(\ref{Sqsp2})), which is a {\it convex} increasing function of
$C(|\Psi\rangle)$. The quantity $\frac{1+\sqrt{1-C^2(\rho)}}{2}$ is also the
maximum fidelity between $\rho$ and a convex mixture of gaussian states, as
shown in \cite{Os.14} with a different treatment based on group-theoretical
methods.

A general mixed state $\rho$ satisfying $[\rho,P]=0$ will be a
convex mixture of pure states with even and odd number parity. It can be
written as a convex mixture of even and odd parts, i.e.,
\begin{equation}\rho=p_+\rho_++p_-\rho_-\,,\label{rhg}\end{equation}
where $\rho_\pm=\frac{1}{2 p_{\pm}}(1\pm P)\rho$ are the even and odd
components of $\rho$ and
$p_{\pm}={\rm Tr}\,\rho(1\pm P)/2$ the corresponding probabilities. Since we
just consider pure states with definite number parity,
for the general mixed states (\ref{rhg}) we may just take
 $E^{\rm qsp}(\rho)=p_+E^{\rm qsp}(\rho_+)+p_-E^{\rm qsp}(\rho_-)$,
with  $E^{\rm qsp}(\rho_{\pm})$ evaluated with Eqs.\ (\ref{Cr}) and (\ref{Er}).

As illustration, we consider a definite parity mixture of a maximally
entangled state $|\Psi\rangle$
($C(|\Psi\rangle)=1$)  with the fully mixed state,
\begin{equation}\rho=p|\Psi\rangle\langle\Psi|+(1-p)I_8/8\,,\end{equation}
where $0\leq p\leq 1$. In the odd parity case, $|\Psi\rangle$ can be written
in the form
\begin{equation}|\Psi\rangle={\textstyle\frac{1}{\sqrt{2}}}(c^\dagger_1|0
\rangle+c_1|\bar{0}\rangle)=
{\textstyle\frac{1}{\sqrt{2}}}(c^\dagger_1+c^\dagger_2 c^\dagger_3 c^\dagger_4
)|0\rangle\,,\end{equation}
whereas in the even parity case we can take $|\Psi\rangle=
\frac{1}{\sqrt{2}}(|0\rangle+|\bar{0}\rangle)$ or
$\frac{1}{\sqrt{2}}(c^\dagger_1c^\dagger_2+c^\dagger_3c^\dagger_4)|0\rangle$.
A direct calculation using (\ref{Cr}) leads to
\begin{equation}C(\rho)={\rm Max}[{\textstyle\frac{7p-3}{4}},0]\,,\end{equation}
indicating entanglement for $p>3/7$, i.e.  $q>1/2$,  where $q=\langle \Psi|
\rho|\Psi\rangle=p+(1-p)/8$ is the total weight of $|\Psi\rangle$.
A similar calculation but considering just two-fermion states,
 $\rho_2=p|\Psi\rangle\langle\Psi|+(1-p)I_6/6$,  leads instead to
$C(\rho_2)={\rm Max}[\frac{5p-2}{3},0]$,
implying entanglement above a slightly smaller
value of $p$ ($p>2/5$, entailing again $q=p+(1-p)/6>1/2$), with
$C(\rho_2)>C(\rho)$ for $p\in(2/5,1)$.
As in the two-qubit case, the existence of a finite threshold probability $p$
for non-zero $C$ and hence $E^{\rm qsp}$, implies a finite limit temperature
for entanglement $T_L$ if $\rho$ represents a thermal state
($\frac{q}{(1-p)/8}\propto e^{-\beta(E_0-E_1)}$,
with $E_0$ the energy of $|\Psi\rangle$ and $E_1>E_0$ that of remaining $
7$ levels), which is larger in the second canonical case.
\vspace*{-.25cm}

\section{Conclusions}

We have presented a general consistent formalism for describing
entanglement-like correlations in general fermion states with no definite
fermion number yet fixed number parity. We have first defined a  single level
entanglement entropy  that quantifies the entanglement between a
single-particle mode and its orthogonal complement, through the definition of
suitable reduced states for such a partition of a given basis of the
single-particle space. The sum over all sp modes of this entropy, $S_{\bm c}$,
can be taken as a measure of the total entanglement of the system with respect
to this basis, and its  minimum over all sp bases, $S^{\rm sp}$, was shown to
be a function of the one-body density matrix, being then invariant with respect
to unitary transformations in the single-particle space. Moreover, if
minimization is extended over all quasiparticle basis, the resulting
entanglement entropy, $S^{\rm qsp}$, is a function of the generalized one-body
density matrix, remaining therefore invariant under general Bogoliubov
transformations. Such entropy vanishes iff there is a single particle or
quasiparticle basis in which every level is separable from its orthogonal
complement, i.e., iff each of these levels is either empty or occupied. These
entanglement entropies satisfy the inequality chain $S_{\bm c}\geq S^{\rm
sp}\geq S^{\rm qsp}$. The convex roof extension of $S^{\rm qsp}$ was also
introduced, its vanishing rigorously identifying ``classically'' correlated
mixed fermion states which can be expressed as convex mixtures of pure states
or quasiparticle vacua, like those emerging at sufficiently high temperatures
in interacting many-fermion systems through approaches like the SPA.

In the case of fermion systems with four single particle levels, a fermionic
analog of the two-qubit pure state concurrence was defined in terms of
$\rho^{\rm qsp}$, which reduces to the Slater correlation measure defined in
\cite{Sch.01,Eck.02} for two- fermion states. The eigenvalues of the
generalized one-body density matrix, which are four-fold degenerate, can be
written as functions of this concurrence and consequently, the entanglement
entropy $S^{\rm qsp}$ is related to the fermionic concurrence by an expression
analogous to that of the two-qubit case. This result suggests that ``particle
entanglement'' may be seen as a minimum ``mode entanglement''. For mixed states
with fixed number parity of this system,  an explicit expression for the
fermionic concurrence, defined as the convex roof extension of the pure state
concurrence, was derived, in complete analogy to the two-qubit case, which
generalizes the result of \cite{Sch.01,Eck.02} and provides a closed analytic
expression for the convex roof extension of $S^{\rm qsp}$.

\acknowledgments
The authors acknowledge support from CONICET (NG) and CIC (RR) of Argentina.
\\

\appendix
\section{Quasiparticle vacuum}
According to Thouless theorem \cite{Th.60} the vacuum $|0_{\bm a}\rangle$ of
the quasiparticle fermion operators
(\ref{anu}) is given, if ${\rm Det}\, U\neq 0$, by \cite{RS.80}
\begin{eqnarray}|0_{\bm a}\rangle&=&\gamma\,{\textstyle\exp[\frac{1}{2}
\sum_{i,j}T_{ij}c^\dagger_i c^\dagger_j]|0\rangle} \nonumber\\
&=&\gamma\,[1+{\textstyle\frac{1}{2}\sum_{i,j}T_{ij}c^\dagger_i
 c^\dagger_j+\ldots]|0\rangle}\label{0a}\,,\end{eqnarray}
where $\gamma=\sqrt{|{\rm Det}\,U|}$ and $T=-U^{-1}V$ is an antisymmetric
matrix, with $|0\rangle$  the vacuum of
the $c_j$ operators. Eq.\ (\ref{0a}) can be verified by directly applying
$a_\nu$ to (\ref{0a}) (if ${\rm Det}\, U=0$, $|0_a\rangle$ can be obtained by
applying additional creation operators $c^\dagger_j$ to Eq.\ (\ref{0a})).

If $|\Psi\rangle=|0_{\bm a}\rangle$, then $f_\nu=\langle 0_{\bm a}|a^\dagger_\nu
a_\nu|0_{\bm a}\rangle=0$ $\forall$ $\nu$, implying $S^{\rm qsp}=0$. However, it is
easy to see that
\begin{equation}
\rho^{\rm sp}=1-\langle 0_{\bm a}|\bm{c}\bm{c}^\dagger|0_{\bm a}\rangle=VV^\dagger
\,,\end{equation}
implying  $S^{\rm sp}>0$ if $V\neq 0$. The eigenvalues $p_k$ of $\rho^{\rm sp}$
are  then just  the square of the singular values of $V$. The state
$|\Psi\rangle$ appears, therefore, mixed at the sp level, reflecting that it
cannot be written as a Slater determinant in operators of the form (\ref{uca}).

\end{document}